\documentclass[preprint,12pt,3p]{elsarticle}

\usepackage[T1]{fontenc}
\usepackage{textcomp}
\usepackage{color,soul}
\usepackage{gensymb}
\usepackage{verbatim}
\usepackage{paralist}
\usepackage{nicefrac}

\usepackage{amsthm}
\usepackage{amsmath}
\usepackage{amsfonts}
\usepackage{amssymb}

\usepackage{array}
\usepackage{url}
\usepackage{color,soul}
\usepackage{cleveref}
\usepackage{subcaption}
\usepackage{mwe}
\usepackage{todonotes}
\usepackage{etoolbox}

\theoremstyle{definition}

\newtheorem{Assum}{Assumption}[section]

\newcommand{\RN}[1]{%
  \textup{\uppercase\expandafter{\romannumeral#1}}%
}

\newcommand{\SGRE}{\textit{Siemens Gamesa Renewable Energy} }
\newcommand{\MADE}{\textit{Manufacturing Academy of Denmark} \text(MADE) }
\newcommand{\IB}{\textit{IntegralBlades}$^\text{\textregistered}$ }
\definecolor{darkblue}{rgb}{0,0,0.5}
\definecolor{brown}{rgb}{0.65,0.10,0.1}

\newtheorem{remark}{Remark}[section]

\begin{document}

\begin{frontmatter}

\title{Stochastic Modelling of the Flow-Front Evolution in a Vacuum Assisted Resin Transfer Moulding Process with Missing Data}

\author[label1,label3]{Michael Nauheimer}
\ead{Michael.Nauheimer@siemensgamesa.com}

\author[label2,label3]{Rishi Relan \corref{cor1}}
\ead{risre@dtu.dk}

\address[label1]{Siemens Gamesa Renewable Energy A/S, Aalborg, Denmark}
\address[label2]{DTU Compute, Technical University of Denmark, Kongens Lyngby, Denmark}
\address[label4]{Centre for Mathematical Sciences, Lund University, Lund, Sweden}

\cortext[cor1]{Corresponding author}
\fntext[label3]{Equal Contributions}

\author[label2]{Uffe H{\o}gsbro Thygesen}
\author[label4]{Erik Lindstr{\"o}m}
\author[label2]{Henrik Madsen}

\begin{abstract}
The real-time fault monitoring and control of the Vacuum Assisted Resin Transfer Moulding (VARTM) production process requires a knowledge of the position of the epoxy flow-front inside the mould. Therefore, a fast and accurate flow-front tracking system capable of combining the underlying physics of the flow-front dynamics with the measured data is highly prized. Stochastic differential equations (SDEs) based grey-box models deliver a good trade-off between high fidelity models and data-driven black-box models for designing such a flow-front position tracking system. In this paper, we propose a simple yet novel coupled SDE based spatiotemporal grey-box model of the flow-front dynamics in case of missing sensor information. The proposed method uses the finite difference approximation of the spatial domain of the flow-front for estimating spatial flow pattern of the epoxy. Furthermore, to accommodate for the missing sensor data, we utilise a modified version of the continuous-discrete extended Kalman filter (CD-EKF) based estimation framework for SDEs that takes into consideration the effective dimension of the measurement space during the identification process. The performance of the method is evaluated for various common sensor faults scenarios at different levels of measurement noise and sampling rates. 
\end{abstract}


\end{frontmatter}

\section{Introduction}
The growing demand for the integration of renewable energy sources into grid has contributed to a surge in the wind power applications. This suggests a move towards larger wind turbines and thereby larger turbine blades. The Vacuum Assisted Resin Transfer Moulding (VARTM) process is used to cast a large-scale composite shell structure like wind turbine blades and aerospace structures \citep{grimsley2003a, song2004modeling, grimsley2004preform}. However, due to an increased blade size and the inhomogeneous nature of the flow inside the mould, the risk of moulding defects such as dry spots and voids increases. This leads to deterioration of the mechanical properties of the cast parts and aggravates the risk of structural failures prompted by errors during the casting of the blades \citep{SREEKUMAR2007453, PARK2011658, MATUZAKI2015154}. 

Ensuring the quality of the cast blades requires a good knowledge about the epoxy flow inside the mould. The patented \IB production method based on the VARTM process used at \SGRE has no possibility of real-time visual inspection of flow-front evolution during the casting process \citep{stiesdal2006method}. Hence, real-time tracking and localization of the flow-front is needed to detect the potential moulding defects such as voids and dry spots. This is because the fault location detected in real-time can be different from the position when the blade is inspected after casting. A better real-time knowledge of the flow-front paves the way for designing an automated multi-inlet production process for larger blades by controlling the opening/closing of the inlets and flow-rate inside the mould. Furthermore, a controlled trajectory of the flow-front inside the mould decreases the risk of areas with dry glass, which eventually decreases the repair time and increase the general quality of the produced blade.

Development of sensor technologies for real-time monitoring of the VARTM processes is an active field of research. Several types of sensors have previously been implemented in VARTM processes including permittivity sensors \citep{YENILMEZ2009476}, pressure sensors \citep{ZHANG20111478}, and sensors based on electrical time-domain reflectometry \citep{DOMINAUSKAS200367}. In general these sensors are accurate but limited to measuring on or close to the surface of the moulded parts. It has been shown in the past that sensors based on optical fibres \citep{KUEH2002311} can be cast into parts making it possible to measure through thickness of the shell. Recently,  \citep{MATSUZAKI201643} proposed a method using two-sided visual observations for 4D data assimilation to accurately reconstruct the 3D resin flow and permeability filed of a fibre preform. 

Furthermore, most of the sensors modalities reported in the literature either rely on visual observations or are based on sensors being in direct contact or close proximity to the epoxy resin. Especially in a hostile environment with harsh chemicals and temperatures reaching up to $200^\circ$ C, the risk of sensor faults and failures increases. In case of a sensor-only setup, a sensor fault or failure may affect the ability to properly measure the progression of the flow-front. Data-driven estimation of spatio-temporal and stochastic differential equation (SDE) based models  are extensively applied to problems in ecological, Geo-statistical or financial statistics fields \citep{dewar2009data,cressie2011statistics}. Typically these are non-parametric models  that do not consider the underlying physics of the system. Hence, such models are not suitable for estimating the flow-front dynamics, where the general physics is known but process parameters vary depending on time and the location of the flow-front inside the mould. 

A first attempt to model the flow-front evolution for a virtual sensing system was made in \citep{nauheimer2018estimation,nauheimer2018stochastic}. In this approach the spatiotemporal evolution of flow-front is modelled by discretising the spatial domain into equidistant cells using an arrangement of parallel line sensors. The authors then utilised the coupled stochastic differential equations (SDEs) to estimate the flow of the epoxy along these multiple flow lines. A $2^{\textup{nd}}$-order finite difference spatial approximation along these line sensors was used to capture the local spatial pattern. However, the  authors did not discuss the validity of the proposed approach in case of missing or faulty information from failing line sensors. %The proposed SDEs based approach can be easily extended to handle missing information/data. 
Higher-order approximation of PDEs are commonly used in many applications e.g. modelling the low-density polyethylene tubular reactors in \citep{zavala2009optimization}. Hence, in this paper, we propose to use the higher-order approximation of the spatial domain of the fluid flow \citep{hirsh1975a} along with a modified version of continuous-discrete extended Kalman filter  (CD-EKF) \citep{KRISTENSEN2004225} based estimation framework for SDEs to handle the missing information from the line sensors. The advantage of the modified version is that it takes into consideration the effective dimension of the measurement space during the estimation process.  

The paper is structured as follows: Section \ref{sec:vartm} briefly describes the VARTM process and the setup for simulating the evolution of the flow-front in a porous medium. The coupled-SDEs based grey-box modeling approach for estimating the flow-front dynamics is described in Section \ref{sec:sdeform}. Section \ref{sec:results} describes the results of the simulation based experiments. Finally the conclusions are given in Section \ref{sec:conc}.

\section{The VARTM process}
\label{sec:vartm}
The VARTM process is a special variety of Resin Transfer Moulding (RTM) processes. As the name indicates the process is only assisted by the difference of the near-vacuum inside the mould and the surrounding pressure of approximately one bar. 
\begin{figure}[!ht]
\centering
\includegraphics[width=0.75 \textwidth]{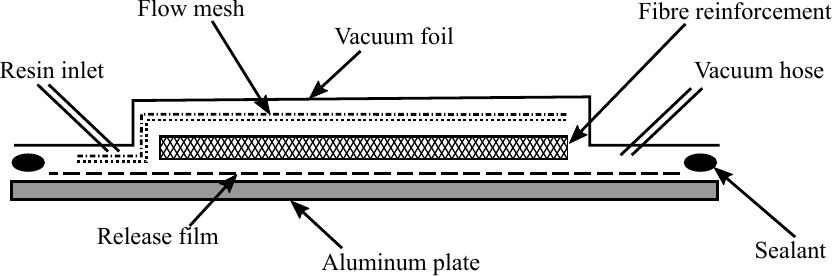}
\caption{The general material configuration and work-flow of a VARTM process. The resin enters the vacuum bag through the resin inlet and spreads into the pores of the fibre reinforcement from left to right. This process is only assisted by the vacuum generated through the vacuum hose.}
\label{fig:vartm}
\end{figure}
In the VARTM process a blade casting mould prepared with glass fibres and other auxiliary materials, is infused with a liquid epoxy driven by the pressure difference  between the inside of the mould of $\approx 0$ bar and the ambient pressure lof $\approx 1$ bar (see Fig. \ref{fig:vartm}). The flow direction and velocity inside the mould is mainly determined by the pressure gradient affected by the individual placement, permeability and porosity of the auxiliary materials inside the mould but also by the local temperatures and their effect on viscosity of the epoxy. 
\begin{figure}[!ht]
\centering
\includegraphics[width=0.65 \textwidth]{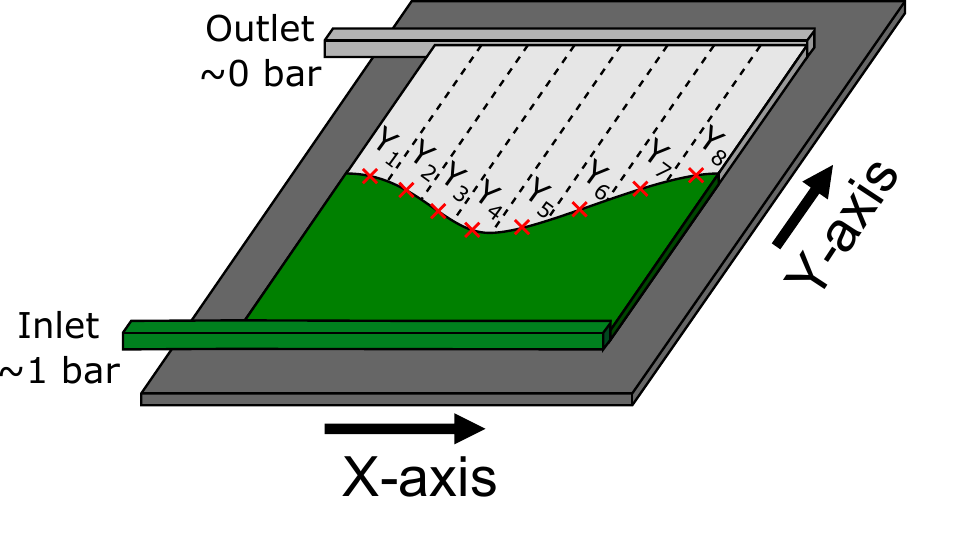}
\caption{Spatially coupled SDEs are used to estimate the spatiotemporal evolution of the flow-front along the $y$ direction. The dotted lines represent the line sensors used to split the spatial domain into equidistant cells across the $x$ direction. The set of line sensors measure the flow along the $y$ direction in the points marked by the red crosses.}
\label{fig:flow_front}
\end{figure} 
The viscosity of the epoxy changes with temperature and continuous mixing of the epoxy with the hardener component during the entire infusion process. %as it starts to cure when epoxy resin is mixed with a hardener component. 
For example, when the epoxy is mixed with the hardener component, it starts to cure immediately, increasing the viscosity over time. However, the continuous mixing of the epoxy during the infusion of the mould decreases the cure rate. In addition, the viscosity of the epoxy is dependent on its temperature, with a higher temperature resulting in a lower viscosity but an increased cure rate. A decreased viscosity will result in an increased flow rate, but the epoxy curing process is an auto-catalytic exothermic process which will result in a fast increase of the temperature and thereby viscosity if heat cannot escape the blade mould. However, for large scale structures, epoxy resin and hardener is usually continuously mixed during the entire infusion process. 
\subsection{Flow in a porous medium:}
In this section, the evolution of the flow of the epoxy inside a rectangular mould (see Fig. \ref{fig:flow_front}) is formulated as a spatio-temporal estimation problem and is described using partial differential equations (PDEs). For the analysis present in this paper, we explicitly state here the following assumptions. 
\begin{Assum}
A PDE model based on Darcy's law in two spatial dimensions provides a good description of the physics of the flow in three spatial dimensions. 
\end{Assum}
\begin{remark}
The thickness of the laminate is generally much smaller than the other dimensions, hence $2$-dimensional formulation can provide a reasonably good approximation. However, the analysis can be extended to $3$-dimensional formulation of the problem.
\end{remark}
\begin{Assum}
Epoxy is considered a Newtonian fluid with constant viscosity.
\end{Assum}
\begin{remark}
Uncured epoxy is not a Newtonian fluid and the viscosity increases continuously as the epoxy cures with time. However, it is expected that minimal or equally distributed force will be exerted on the epoxy within the mould and that the increase in viscosity caused by increasing cure degree is low for short periods of time. Thereby it can be considered a Newtonian fluid with constant viscosity.
\end{remark}
\begin{Assum}
Epoxy enters the mould evenly across a line perpendicular to the direction of the flow.
\end{Assum}
\begin{remark}
In an experimental test setup or full scale casting the epoxy enters the mould through one or several inlet points. However, the epoxy typically flows in a cavity which ensures a homogeneous inlet across a line perpendicular to the flow direction.
\end{remark}

\subsection{Mathematical Formulation of the Flow-Front Evolution}
Darcy's law \citep{darcy1856fontaines} describes the volumetric flow velocity, $\mathbf{q}=\mathbf{q}(x,y,z,t)$ [m$^3$/s] of a fluid in a porous medium in three dimensions. To reduce computational complexity when simulating the flow using PDEs to generate measurement data the model is reduced to include only the two spatial dimensions describing the in-plane flow, 
\begin{equation}
	\mathbf{q} = -\frac{\boldsymbol{\kappa} \varphi H}{\mu}\nabla p
    \label{eq:qimp}
\end{equation}
where $\mathbf{q}=\mathbf{q}(x,y,t)$ is the flow velocity [m/s] integrated over the vertical axis, $\boldsymbol{\kappa}=\boldsymbol{\kappa}(x,y)$ is the permeability tensor for the porous medium [m$^2$], $\varphi$ is the porosity of the medium [-], $H$ is the cross-sectional thickness [m], $\mu=\mu(x,y,t)$ is the fluid viscosity [Pa$\cdot$s], $p=p(x,y,t)$ is the pressure [Pa], and $\nabla=(\frac{\partial}{\partial_x},\frac{\partial}{\partial_y})$ is the in-plane spatial derivative [1/m]. In combination with the conservation of mass this results in
\begin{equation}
	\dot{h}+\nabla\cdot \mathbf{q} = 0
	\label{eq:mass_cons}
\end{equation}
where $h=h(x,y,t)\leq H$ is the thickness of the fluid layer. By assuming the following relationship 
\begin{equation}
	h = \text{min}\left(\varphi H,\frac{p}{\rho g}\right)
	\label{eq:thickness}
\end{equation}
where $g$ is the gravitational acceleration and $\rho$ is the density of the fluid. This corresponds to local hydrostatic equilibrium, in that the pressure is proportional to the thickness of the fluid layer where the gap is partially filled with fluid, but may be larger when the gap is completely filled with fluid. This allows us to eliminate $h$ from the model by combining equations \eqref{eq:mass_cons} and \eqref{eq:thickness} to obtain:
\begin{equation}
	\dot{h} = \frac{\text{d}h}{\text{d}p}\dot{p} = \nabla\cdot\left(\frac{\boldsymbol{\kappa} \varphi H}{\mu}\nabla p\right)
	\label{eq:pde}
\end{equation}
The PDE governing $p$ is completed with boundary conditions that assumes no-flux boundary conditions along the sides, a pressure of $p_{0}=1$ bar at the inlet, and a pressure of zero bar at the outlet. 

\begin{figure}[!ht]
\centering
\includegraphics[width=0.65 \textwidth]{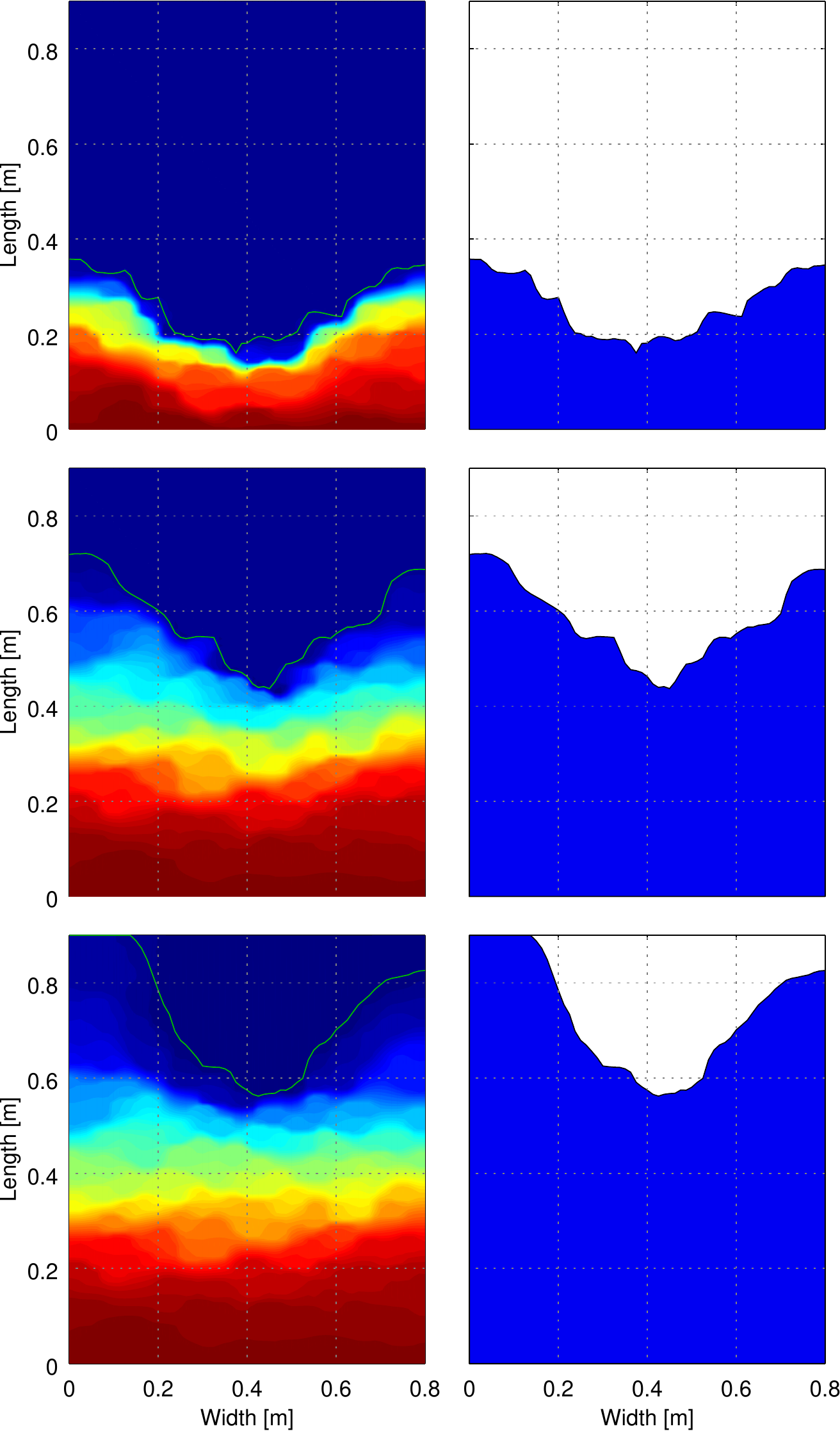}
\caption{The left three plots shows a contour plot of the pressure distribution. The red colour indicates a high pressure of $1$ bar and the blue colour indicates a low pressure of $0$ bar. The green line shows a plot of the generated flow-front measurement data. The right three plots shows the evolution of the generated flow-front measurement data in correspondance to the green line in the plots to the left.}
\label{fig:pressure_flow}
\end{figure}

\subsection{Generation of the Simulated Flow-Front Data}
Simulated flow-front evolution data is generated using the numerical PDE solver \textit{FEniCS Project} \citep{AlnaesBlechta2015a}, which employs the finite element method to discretize space. Output from the model is considered synthetic measurement data for this case-study. The PDE described in \eqref{eq:pde} is solved for a $L_{x}\times L_{y}$ ($80$ cm $\times$ $90$ cm) rectangular system, numerically described by $(n_{x}+1)\times(n_{y}+1)$ (i.e. $65\times 129$) vertices, similar to the one shown in Fig. \ref{fig:flow_front}. 

While specifying the placement of glass fibre and auxiliary materials in a blade mould, normally it is ensured that the evolution of the flow-front during infusion process is homogeneous. However, sometimes due to manufacturing errors materials are handled and placed less carefully than specified, resulting in perturbations in the flow-front progression during the infusion process. Therefore, here a heterogeneous case is simulated. In the heterogeneous case $\frac{\kappa}{\mu}$ is kept constant with respect to (w.r.t) the time but along the spatial directions we choose to model it as
\begin{equation}
	\frac{\kappa}{\mu} = \frac{c_{0}}{\left(1-A\cdot \cos\left(\frac{2\pi x}{L_{x}}\right) \right)\left(1-A\cdot\cos\left(\frac{2\pi y}{L_{y}}\right)\right)}
    \label{eq:111}
\end{equation}
where $A$ is a constant determining the relative decrease in permeability towards the middle, $c_{0}$ being a correction constant and $L_{x}=0.8$ m ; $L_{y}=0.9$ m are the width and length of the casted area. This model describes a decrease of the permeability towards the centre of the cast area.  
\begin{remark}
Any nonlinear function can be used here to simulate the change in permeability along the spatial domain.
\end{remark}
The numerical solution is obtained by using a semi-implicit Euler solver to time-march, where the derivative $\dot{p}$ is approximated by a first-order finite difference and the derivative $\frac{dh}{dp}$ is evaluated at the previous time step. The right-hand side of \eqref{eq:pde} is evaluated at the next time step. The system can be seen as differential-algebraic, since the derivative $\frac{dh}{dp}$ is zero in those parts of the spatial domain that have already been impregnated with the fluid. The simulated flow-front evolution data $\mathcal{Z}_{l,t}$ for each line, $l=1,...,n_{x}+1$, is generated by evaluating which of the $n_{y}+1$ vertices along the $y$-axis, where the pressure is above a threshold, $p_{\text{th}}$, for all $n_{x}+1$ vertices spread across the $x$-axis of the system.
\begin{equation}
    \mathcal{Z}_{i,t} = \frac{\sum\limits_{\mathbb{K}=1}^{n_{y}+1}\text{max}(\text{min}(p(x_{n},y_{\mathbb{K}},t),p_{\text{th}}),0)}{p_{\text{th}}\cdot(n_y+1)}L_{y}
    \label{eq:measurements}
\end{equation}

Fig. \ref{fig:pressure_flow} shows a heterogeneous flow-front progression. The left three plots shows a contour plot of the pressure distribution. The red colour indicates a high pressure of $1$ bar and the blue colour indicates a low pressure of $0$ bar. The green line shows a plot of the generated flow-front measurement data. The right three plots shows the evolution of the generated flow-front measurement data in correspondence to the green line in the plots to the left. In Fig. \ref{fig:pressure_flow} it is seen how the decrease in the permeability also causes decreased flow rate towards the middle. 
Although these high-dimensional PDEs based models are good to understand and simulate a dynamical process they are not very useful when the final intended purpose of the model is control and monitoring. Hence in the next section, we propose to use a coupled stochastic differential equation (SDE) \citep{oksendal2010stochastic} based modelling approach to model the flow-front dynamics.

\section{SDE based grey-box modelling}
\label{sec:sdeform}
Grey-box models are typically SDEs based models \citep{KRISTENSEN2004143, kristensen2004method}, where the structure of the model is built based on a combination of physical knowledge of the system, as in the white-box models, and on the statistical information based on the observations (measured data), as in the case of the black-box models. From a theoretical point of view, SDEs are the preferred choice to model stochastic, complex, and nonlinear systems where only a partial information about the system dynamics is available. Here, we consider the grey-box model structure \citep{bohlin1995issues} consisting of a set of nonlinear, discretely, partially observed SDEs with measurement noise. 

In the SDE formulation, the time evolution of the states of a dynamical system is separated into the drift term $f_t$ and the diffusion term $\sigma_t$ respectively. The drift and the diffusion terms can either be represented by a linear or nonlinear function. This separation allows to separate the modeling error into the error due to unmodelled dynamics (i.e. model approximations and noise originating from unknown disturbances to the system) using the diffusion term  and the measurement noise (the uncorrelated error due to imperfect measurement equipment), resulting in an accurate description of the system dynamics \citep{oksendal2010stochastic} as described below
 \begin{align}
    \text{d}Y_t &= \underbrace{f_t(Y_t,U_t,t,\theta)\text{d}t}_\text{Drift} + \underbrace{\sigma_t(Y_t, U_t,t,\theta)\text{d}W_t}_\text{Diffusion}; \hspace{0.5em }t_0 \leq t \leq T,
    \label{eq:SDE1}
\end{align} where $t \in \mathbb{R}$ represents the time variable and $t_k$, $k = 0, \cdots, N$ represents the sampling instants. The  state equations containing the stochastic state variables $Y_t \in \mathcal{Y} \subset \mathbb{R}^n$ of the system are  formulated in continuous time. $Y_{t_0}$ is the stochastic initial condition which satisfies $\mathbb{E}[\left\Vert\left({Y_{t_0}}\rm\right) \right\Vert^2] < \infty$ and $U_t \in \mathbb{R}^d$ is the vector of the system's deterministic inputs which is known for all $t$. In \eqref{eq:SDE1}, to ensure the existence of a strong solution, sufficient regularity conditions are assumed for the drift term $f:[t_0,T]\times \mathbb{R}^n \times \mathbb{R}^d \longmapsto \mathbb{R}^n$  and the diffusion term $\sigma: [t_0,T]\times \mathbb{R}^n \times \mathbb{R}^d \longmapsto \mathbb{R}^{n \times m}$, see \citep{oksendal2010stochastic} for further details. 

The process noise $W_t = (W^1_t,W^2_t,\cdots,W^m_t)^T$ is a standard Wiener process of dimension $\mathbb{R}^m$ with an incremental covariance $Q_t$. Due to the identifiability constraints $Q_t$ is assumed to be an identity matrix. The discrete-time observations $Z_{t_k}$  of the observable states are linked to the continuous-time state equation through the continuous differentiable (with respect to $Y_t$) nonlinear function $g_{t_k}(Y_k,U_k,t_k,\theta)\in \mathbb{R}^{\mathcal{L}}$ in the observation  equation described below (\ref{eq:obs1}). 
\begin{equation}
       Z_{t_k}  = g_{t_k}(Y_{t_k},U_{t_k},t_k,\theta) + \underbrace{e_{t_k}}_\text{Measurement noise}
\label{eq:obs1}  
\end{equation} $Z_{t_k} \in \mathcal{Z} \subset \mathbb{R}^{\mathcal{L}}$ represents the vector of the system's outputs; $\theta \in \Theta \subset \mathbb{R}^p$ contains the vector of parameters of the system and $e_{t_k}$ is a $\mathcal{L}$-dimensional zero mean Gaussian white noise process with covariance $S_{t_k}$. Generally, the stochastic entities $Y_{t_0}, W_t$ and $e_{t_k} \in \mathcal{N}(0, S_{t_k}(U_{t_k}, t_k, \theta))$ are assumed to be mutually independent for all $t$ and $t_k$. The solution to \eqref{eq:SDE1} is a stochastic process where the transition probabilities are provided by the Fokker-Plank equation \citep{klebaner2005introduction}. 

\begin{Assum}
The diffusion term is independent of the state variables i.e.
\begin{equation}
    \text{d}Y_t = \underbrace{f_{t}(Y_{t_k},U_{t_k},t_k,\theta)\text{d}t}_\text{Drift} + \underbrace{\sigma_t(U_{t_k},t_k,\theta)\text{d}W_t}_\text{Diffusion}; \hspace{0.5em }t_0 \leq t \leq T,
\end{equation}
\end{Assum}
\begin{remark}
This makes the parameter estimation more feasible. Moreover, the EKF based estimation framework for SDEs requires transformations that can move (or remove) the state dependence from the diffusion term to the drift term to make the filter approximations sufficiently accurate \citep{BAADSGAARD19971369}. For a restricted class of dynamic systems with such dependencies or level effects, a \textit{Lamperti transformation} may be applied to allow the application of the proposed estimation scheme as shown by \citep{nielsen2001applying, moller2010state}.
\end{remark}
\begin{remark}
The interpretation of the SDEs  may be done either in the sense of \textit{Stratonovich} or in the sense of \textit{It{\^o}}. Due to its martingale property with respect to Brownian motion, absence of spurious drift-term etc., the \textit{It{\^o}} interpretation is considered more suitable for parameter estimation  \citep{kloeden2012numerical}, therefore we adapt the \textit{It{\^o}} interpretation here.
\end{remark}

\begin{remark}
There is no exogenous input $U_t$ in the considered problem. However, the algorithm is presented in its general form for completeness.
\end{remark}

\subsection{Finite difference approximation of the spatial domain}
By rewriting of Darcy's Law described in %\eqref{eq:darcy} and 
\eqref{eq:qimp} the following description of the flow-front, along multiple one-dimensional lines as shown in Fig. \ref{fig:flow_front}, is derived. For small values of $H$, the following relations hold; 
\begin{align}
    \frac{dY}{dt} &= \frac{q(y,t)}{\varphi H} = -\frac{\kappa}{\mu}\cdot\nabla p,\\
    p(x,y,t)      &= p_{0}\cdot\text{max}(0,1-\frac{y}{Y_{t}}),
\end{align} then for each line $i$
\begin{align}
	\frac{\text{d}Y_{i,t}}{\text{d}t} = \frac{\kappa p_{0}}{\mu}\frac{1}{Y_{i,t}},
\end{align}
where $Y_{i,t}$ is the flow-front progression [m/s] along line $i$ where, $i=1,...,n$. These equations assume perfect homogeneity of the flow-front. Hence, to account for the heterogeneous nature of multidimensional flow, a spatial descretization, $\mathcal{G}_{i,j}(Y_{i,t},t)$, is introduced along the $x$-axis together with an active diffusion term $\sigma_{i,t}\text{d}W_{t}$ to parameterize any differences between the model and the true system
\begin{equation}
    \text{d}Y_{i,t} = \left(\frac{C_{0,i}}{Y_{i,t}} + D_{0}\mathcal{G}_{i,j}(Y_{i,t},t)\right)\text{d}t+\sigma_{i,t}\text{d}W_{t}
\end{equation} 
where $D_0$ is the spatial coupling coefficient between two adjacent line sensors, $C_{0,i}$ is the value of $\frac{\kappa p_{0}}{\mu}$ for each line, $i$, and $\mathcal{G}_{i,j}(Y_{i,t},t)$ is a $j^\text{th}$-order finite difference approximation of the spatial domain along the $x$-axis \citep{ alma991001871549703821}. 
The $4^{th}$-order central finite difference approximation $\mathcal{G}_{i,4}$ for one of the line sensors can be written as,
\begin{equation}
\mathcal{G}_{i,4}(Y_{i,t},t) = \frac{({Y_{i-2,t}-4Y_{i-1,t}+6Y_{i,t}-4Y_{i+1,t}+Y_{i+2,t})}}{(\Delta x)^{4}},
\end{equation}
where $\Delta x$ represents the spatial distance between the line sensors along the $x$ direction.

\begin{remark}
Central finite difference approximations are used except for the boundary cases ($i=1,..,\frac{j}{2}$ or $i=n-\frac{j}{2}+1,...,n$) where the symmetric forward or the symmetric backward finite difference approximations are used respectively.
\end{remark}
\begin{remark}
Here, the spatial discretisation describes the number of equidistant cells the spatial domain is split into across the $x$ direction whereas the spatial approximation describes the approximation used in the SDEs based models to estimate the flow-front dynamics, i.e. the number of neighbouring line sensors included in the model for the estimation along the chosen line sensors.
\end{remark}

\subsection{Maximum Likelihood Estimation of SDEs}
There are many methods suggested in the literature for parameter estimation in SDEs \citep{kloeden1989survey, shoji1997comparative, singer2004survey}. In this paper, we formulate the problem as a maximum likelihood estimation problem \citep{KRISTENSEN2004225}. The maximum likelihood method is assumes the normality of the model residual. The parameters, $C_{0,i}$ and $D_{0}$, of the coupled SDEs formulated above to model the evolution of the flow-front dynamics are estimated from the measured data simulated using the complex PDE based model.  Given the sequence of measurements $\mathcal{Z}_N$, the likelihood function is formulated using the one-step prediction errors, $\epsilon_k = z_{t_k} - \hat{z}_{t_{k-1}} $, and the associated variances, $R_{t_k|t_{k-1}} =$ Var$(z_{t_k} | \mathcal{Z}_{t_{k-1}}, \theta)$ as below \citep{KRISTENSEN2004225}:
\begin{align}
     \label{eq:likelihood}
    \mathbb{L}(\theta;\mathcal{Z}_N) &= p(\mathcal{Z}_N|\theta) \\ 
     &= \Bigg( \prod_{k=1}^{N} \underbrace{\frac{\exp\Big(-\frac{1}{2}\epsilon_{t_k}^TR^{-1}_{t_k|t_{k-1}}\epsilon_{t_k}\Big)}{\sqrt{\det(R_{t_k|t_{k-1}})(\sqrt{2 \pi})^{\mathcal{L}}}} }_\text{$\mathcal{T}$}\Bigg)p(z_0|\theta)\nonumber \\
     & = - \frac{1}{2}\sum_{k=1}^N \Big(\epsilon_{t_k}^TR^{-1}_{t_k|t_{k-1}}\epsilon_{t_k} + \log \text{det} R^{-1}_{t_k|t_{k-1}} + \mathcal{L} \log 2\pi \Big)
\end{align}
where $\theta$ is a set of parameters, $\mathcal{Z}_N$ is the set of observations, $\mathcal{L}$ is the dimension of the observation space, and $z_0$ is initial measurement. The parameter estimates are found by minimizing the negative log-likelihood:
\begin{equation}
    \hat{\theta} = \underset{\theta \in \Theta}{\mathrm{argmin}}\big\{(\mathbb{L}(\theta;\mathcal{Z}_N)|z_0)\big\}.
\end{equation}
The corresponding value of the negative log-likelihood is the observed maximum likelihood value for that data set and model.

\subsection{Continuous-Discrete Extended Kalman Filter} The complex structure of SDEs makes the parameter estimation non-trivial except for some simple cases. Whereas, the discrete-time extended Kalman filter (EKF) is extensively used in modelling and predictive control of nonlinear systems \citep{huang2009robust}. Here, a CD-EKF is used to compute the solution for $\epsilon_{t_k}$ and $R_{{t_k}|{t_{k-1}}}$ iteratively for a given set of parameters and initial states. The output prediction equations of the CD-EKF are formulated as:
\begin{align}
\label{eq:output}
    Z_{{t_k}|{t_{k-1}}}  &= g(Y_{{t_k}|{t_{k-1}}},U_{t_k},t_k,\theta) \\
    \label{eq:var}
    R_{{t_k}|{t_{k-1}}} &= CP_{{t_k}|{t_{k-1}}}C^T+ S_{t_k}
\end{align} where $C = \frac{\partial g}{\partial Y_t}\big|_{\hat{Y}_{{t_k}|{t_{k-1}}},U_{t_k}, t_k}$ is the first order Taylor expansion (\textit{the Jacobian}) of $g$ and $P_{{t_k}|{t_{k-1}}}$ is the conditional variance of the one-step prediction. Similarly the Kalman gain can be calculated as 
\begin{equation}
  K_{t_k} = P_{{t_k}|{t_{k-1}}}C^TR^{-1}_{{t_k}|{t_{k-1}}}  
\end{equation} Note that the Kalman gain is proportional to the information $R_{{t_k}|{t_{k-1}}}$ provided by the $k^{th}$ observation. Finally the updated equations i.e. the description of the predicted state trajectory and the information obtained from the $k^{th}$ observation $Z_{t_k}$, are written as:
\begin{align}
\label{eq:update1}
    \hat{Y}_{{t_k}|{t_{k-1}}} &= \hat{Y}_{{t_k}|{t_{k-1}}} + K_{t_k} \epsilon_{t_k} \\
    \label{eq:update2}
    P_{t_k|t_k}       &= P_{{t_k}|{t_{k-1}}} - K_{t_k} R_{{t_k}|{t_{k-1}}}K_{t_k}^T
\end{align} This leads to the following state equations
\begin{align}
    \frac{d \hat{Y}_{t|t_k}}{dt} &= f(\hat{Y}_{t|t_k},U_t,t,\theta)\\
    \frac{d \hat{P}_{t|t_k}}{dt} &= AP_{t|t_k} + P_{t|t_k} A^T + \sigma \sigma^T
\end{align} which are solved for $t \in [t_k, t_{k+1}[$. In the equations above, the following short-hand notation 
\begin{align}
    A &= \frac{\partial f}{\partial Y_t}\big|_{\hat{Y}_{{t_k}|{t_{k-1}}},U_{t_k}, t_k}, \\
\sigma &= \sigma(U_{t_k}, t_k, \theta), \hspace{0.5em} S_{t_k}= S(U_{t_k}, t_k, \theta)
\end{align} has been applied. The initial conditions $\hat{Y}_{t|t_0} = Y_0$ for the CD-EKF can either be pre-specified by the user or can be estimated as unknown parameters in the overall optimization problem. Similarly $P_{t|t_0}= P_0$ can be computed as the integral of the Wiener process and the system dynamics evaluated over the first sample and scaled by a pre-specified scaling factor $P_s \geq 1$ as 
\begin{equation}
    P_0 = P_s \int^{t_1}_{t_0} e^{AS} \sigma \sigma^T (e^{AS})^T ds.
\end{equation}

\subsection{Missing data handling in the estimation framework}
The missing values in the output vector $Z_{t_k}$ can be handled easily within the existing estimation scheme described in the section above by slightly modifying the term $\mathcal{T}$ in \eqref{eq:likelihood}. The common way to account for the missing observations or in other sense the non-informative data in the CD-EKF estimation framework, is by setting the corresponding element of the covariance matrix $S_{t_k}$ in \eqref{eq:var} to $\infty$, which is equivalent to zeros in the corresponding elements of the matrix $R_{{t_k}|{t_{k-1}}}^{-1}$ as well as the Kalman gain matrix $K_{t_k}$. This implies that the equations \eqref{eq:update1} and \eqref{eq:update2} will not be updated for the missing values. But this particular approach for calculating $\mathcal{T}$ can not be used straightaway, as a solution is needed, which reflects the effective reduced dimension of $Z_{t_k}$ due to the missing values, in the modified $\epsilon_{t_k}$ and $R_{{t_k}|{t_{k-1}}}$. Alternatively, \eqref{eq:obs1} can be replaced by 
\begin{equation}
   \Bar{Z}_{t_k} = \mathcal{P}(h(Y_{t_k},U_{t_k},t_k,\theta) + e_{t_k}) 
\end{equation}
Here $\mathcal{P}$ represents an appropriate permutation matrix. One of the ways to construct $\mathcal{P}$ is by eliminating the rows corresponding to the missing values in $Z_{t_k}$ from a unit matrix. Equivalently, the output prediction equations of the CD-EKF are replaced with the alternative representation as below:
\begin{align}
    \label{eq:outputNew}
    \Bar{Z}_{{t_k}|{t_{k-1}}}  &= \mathcal{P}\hspace{0.1em}g(Y_{{t_k}|{t_{k-1}}},U_{t_k},t_k,\theta) \\
    \label{eq:varNew}
    \Bar{R}_{{t_k}|{t_{k-1}}} &= \mathcal{P}CP_{{t_k}|{t_{k-1}}}C^T\mathcal{P}^T+ \mathcal{P}S_{t_k}\mathcal{P}^T
\end{align} the innovation equation $ \Bar{\epsilon}_{t_k} = \Bar{Z}_{t_k} - \hat{\Bar{Z}}_{{t_k}|{t_{k-1}}}, $ the Kalman gain equation becomes 
\begin{equation}
            \Bar{K}_{t_k} =  P_{{t_k}|{t_{k-1}}}C^T\mathcal{P}^T \bar{R}^{-1}_{{t_k}|{t_{k-1}}}
\end{equation} and finally the update equations are reformulated as,
\begin{align}
    \label{eq:update1New}
    \hat{Y}_{t_k|t_k} &= \hat{Y}_{{t_k}|{t_{k-1}}} + \Bar{K}_{t_k} \Bar{\epsilon}_{t_k} \\
    \label{eq:update2New}
    P_{t_k|t_k}       &= P_{{t_k}|{t_{k-1}}} - \Bar{K}_{t_k} \Bar{R}_{{t_k}|{t_{k-1}}}\Bar{K}_{t_k}^T
\end{align} whereas the state prediction equations remain unchanged, which in turn leads to the modified term $\mathcal{T}^*$:
\begin{equation}
   \mathcal{T}^* =  \frac{\exp\Big(-\frac{1}{2}\Bar{\epsilon}_{t_k}^T\Bar{R}^{-1}_{{t_k}|{t_{k-1}}}\Bar{\epsilon}_{t_k}\Big)}{\sqrt{\det(\Bar{R}_{{t_k}|{t_{k-1}}})(\sqrt{2 \pi})^{\mathcal{\Bar{L}}}}}
\end{equation} where $\mathcal{\Bar{L}}$ is the reduced dimension of the observation space. 
\begin{remark}
For more complex and larger blade geometries with many line sensors and nonlinear flow-front profiles, this is a computationally efficient way to handle missing observations during parameter estimation because the flow-front models may need updating for different blade sizes.
\end{remark}

\section{Simulations, results and discussion}
\label{sec:results}
\begin{figure}[!ht]
    \centering
    \begin{subfigure}[b]{0.45\textwidth} 
        \centering
        \includegraphics[width=0.75 \textwidth]{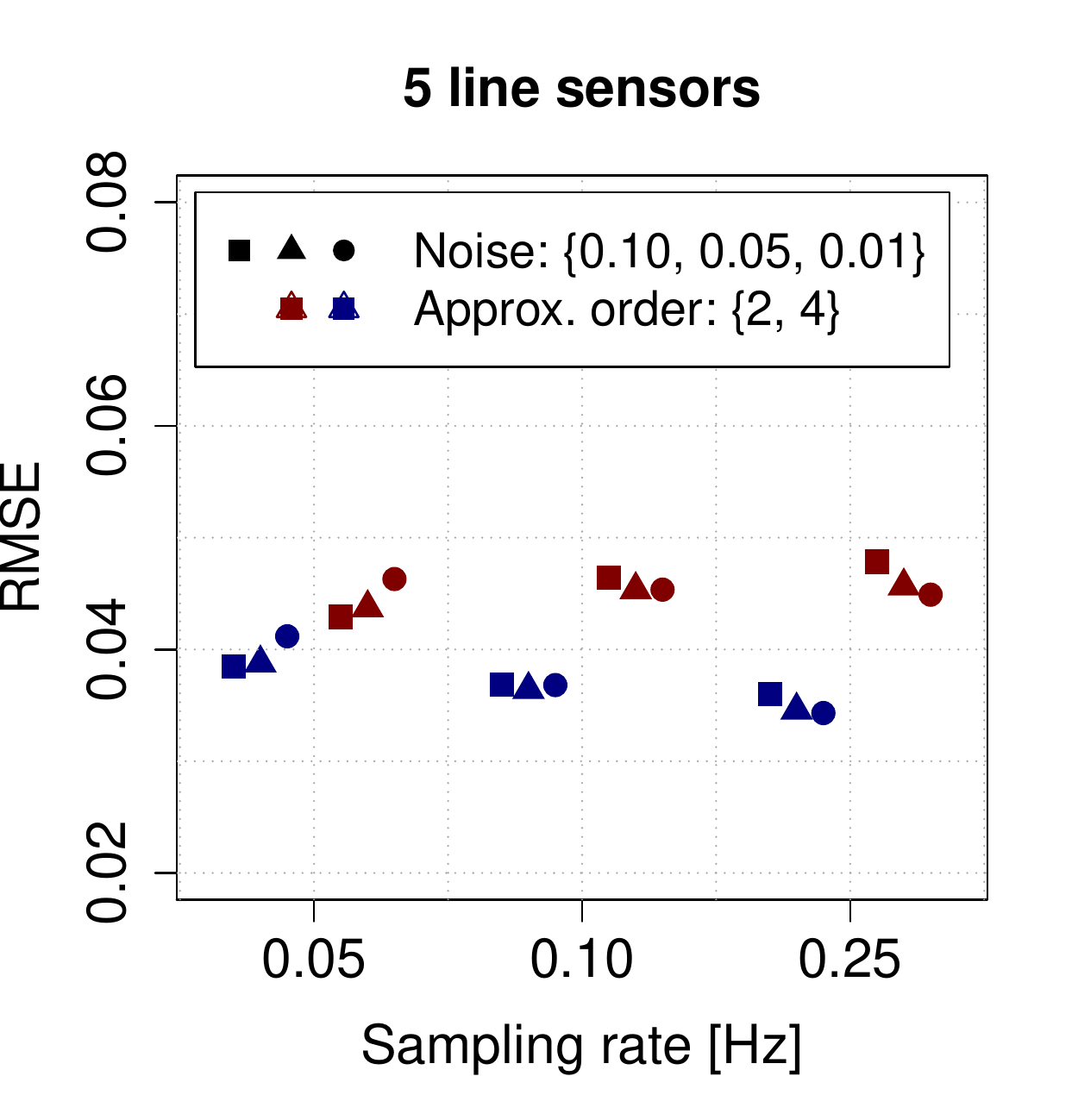}
        \caption{}%\textcolor{blue}{The RMSE values for the predicted flow front using the model fitted to line sensor data from five line sensors. }}
        \label{fig:data_table_plot9}
    \end{subfigure}
    \quad
    \begin{subfigure}[b]{0.45\textwidth} 
        \centering
        \includegraphics[width=0.75 \textwidth]{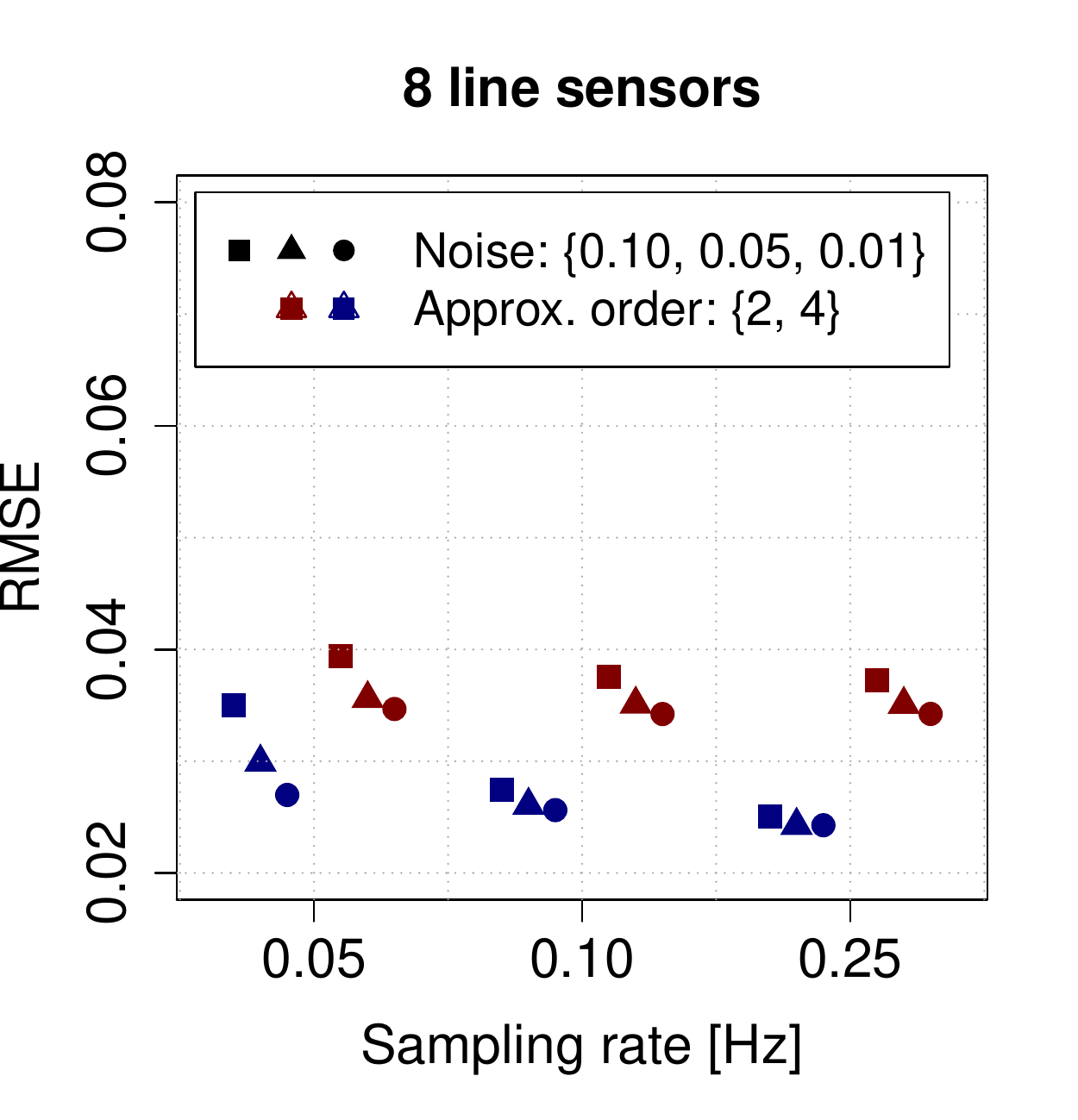}
        \caption{}%\textcolor{blue}{The RMSE values for the predicted flow front using the model fitted to line sensor data from eight line sensors. }}
        \label{fig:data_table_plot10}
    \end{subfigure}
    \vskip\baselineskip
    \begin{subfigure}[b]{0.45\textwidth} 
        \centering
        \includegraphics[width=0.75 \textwidth]{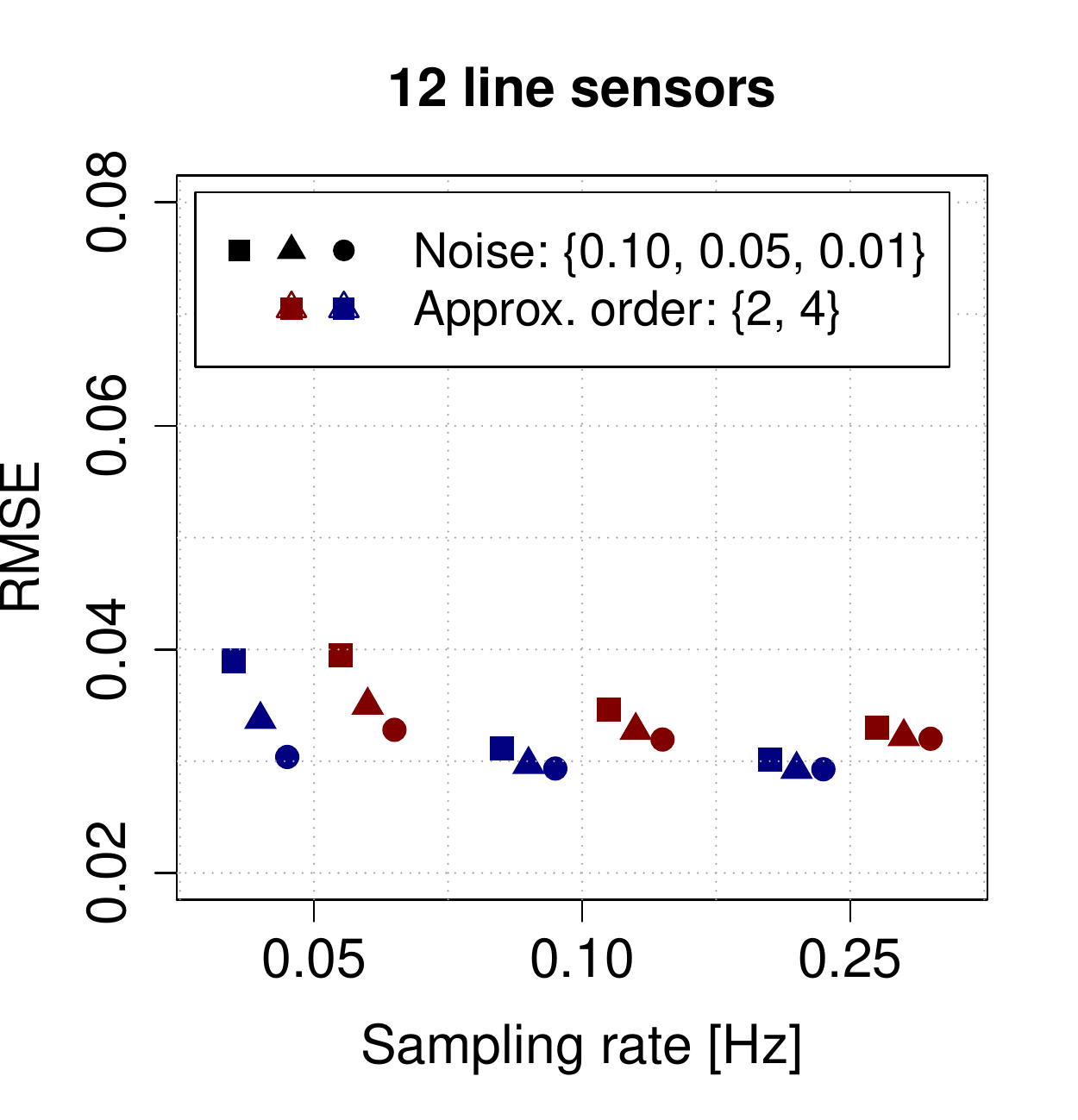}
        \caption{}%\textcolor{blue}{The RMSE values for the predicted flow front using the model fitted to line sensor data from twelve line sensors. }}
        \label{fig:data_table_plot11}
    \end{subfigure}
    \quad
    \begin{subfigure}[b]{0.45\textwidth} 
        \centering
        \includegraphics[width=0.75 \textwidth]{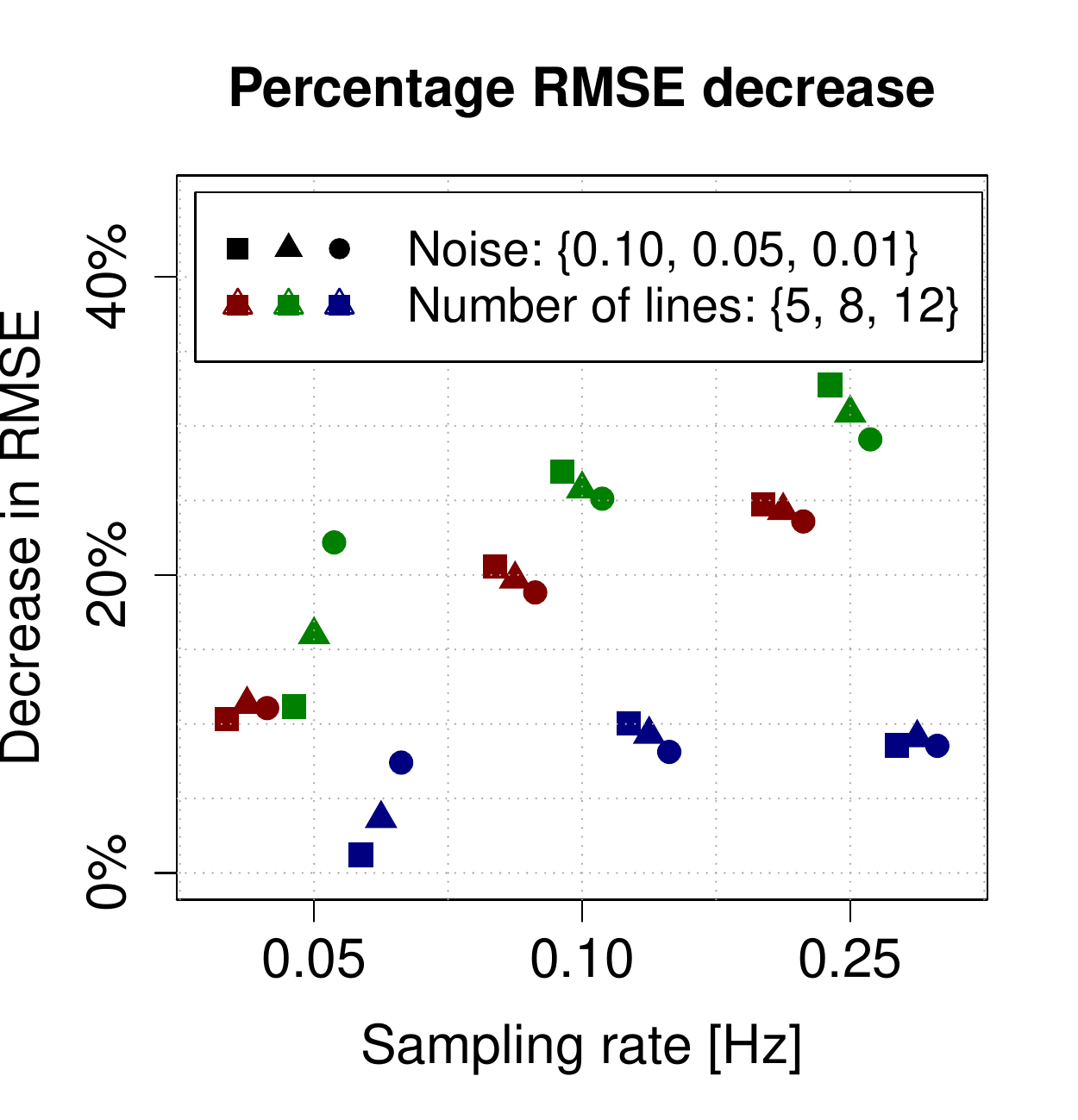}
        \caption{}
        \label{fig:data_table_plot12}
    \end{subfigure}
    \caption{The average RMSE values for the one-step ahead predictions of the flow-front using the models fitted to line sensor data for five, eight and twelve line sensors are shown in (a)-(c). The relative decrease of the average RMSE values when increasing from $2^{\textup{nd}}$-order to the $4^{\textup{th}}$-order finite difference approximation of the spatial domain are shown in (d). The data is perturbed around the specified sampling frequencies for presentation.}
    \label{fig:data_table_plot912}
\end{figure}
This section presents the different validation results for the modeling approach described in the previous section. Firstly, the estimation accuracy is compared for two models of different order of finite difference approximation of the spatial domain. Secondly, a comparison of the estimation accuracy for the two models utilizing $8$ sensors is done for three case studies of missing or faulty sensor information. 
\subsubsection*{Approximation order}
\begin{figure}[!ht]
\centering
\begin{subfigure}[b]{0.45\textwidth} 
\centering
\includegraphics[width= 0.75\textwidth]{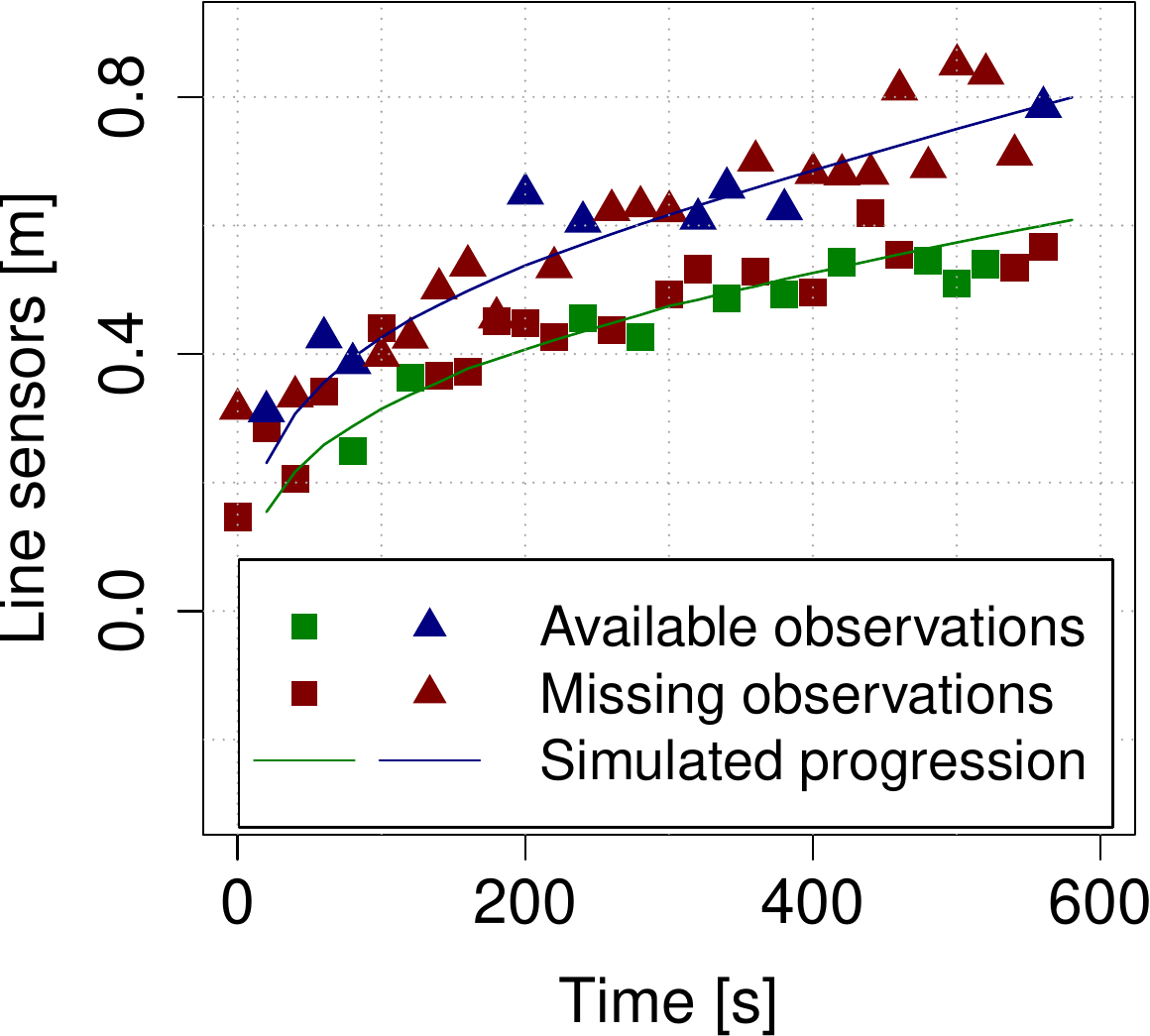}
\caption{}
\label{fig:sensor3_7_miss}
\end{subfigure}
\quad%\vskip\baselineskip
\begin{subfigure}[b]{0.45\textwidth} 
\centering
\includegraphics[width= 0.75\textwidth]{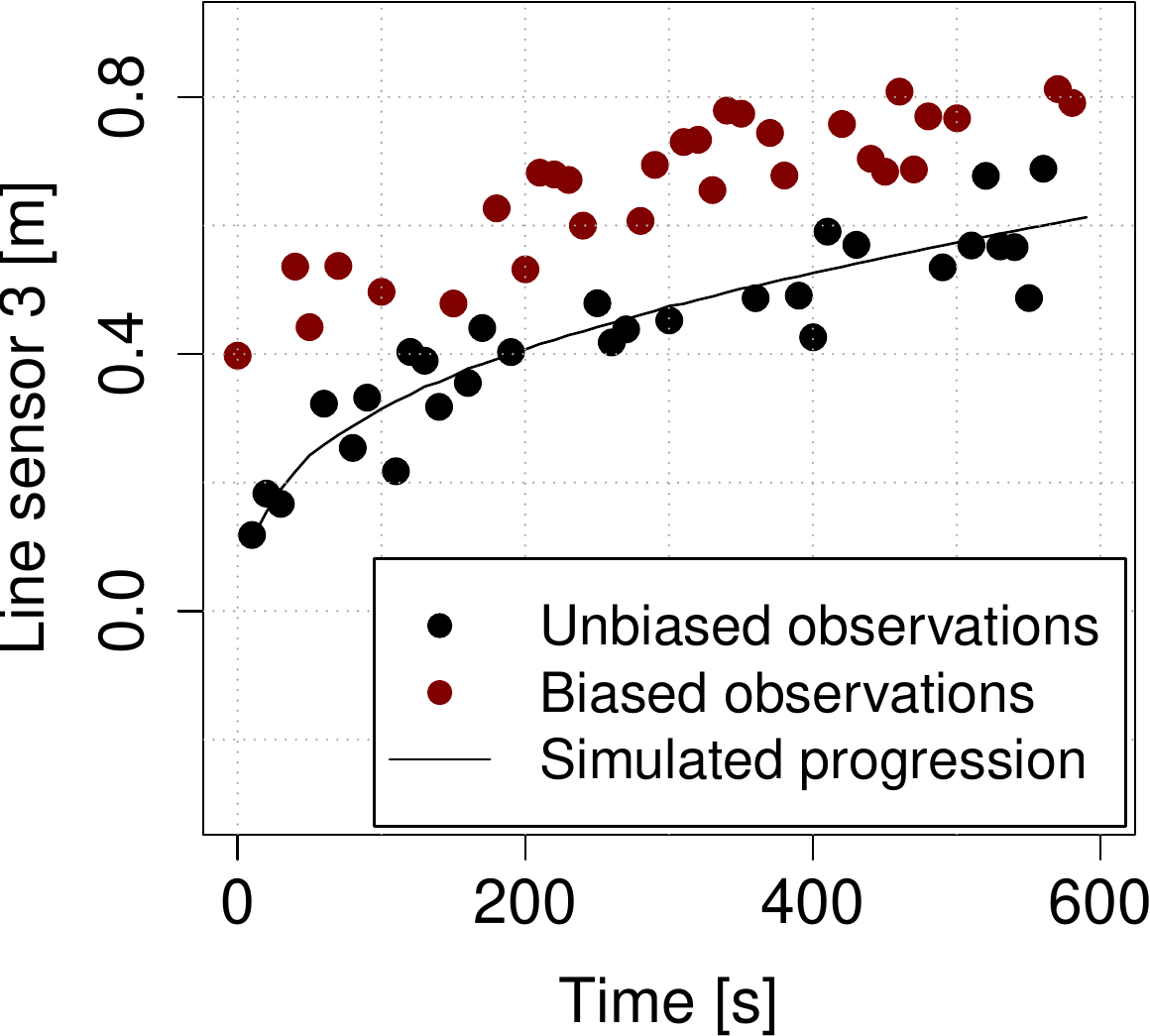}
\caption{}
\label{fig:sensor3_bias}
\end{subfigure}
%\vspace{-0.9cm}
\caption{Shows the plots of the sensor line data with errors. In (a) the simulated case of missing 70\% measurements for line sensors 3 and 7 is shown. In (b) the simulated case of bias on 50\% of the measurement data is shown.}
\label{fig:sensor3_bias_f}
\end{figure}
The comparison is done for two different SDE models i.e. a $2^{\textup{nd}}$-order and a $4^{\textup{th}}$-order finite difference approximation of the spatial domain, respectively. For each time instance the root mean squared error (RMSE) between the simulated flow-front and the one-step ahead predictions of the flow-front is calculated. The average of the calculated RMSE values across all time instances is considered to be a measure of the estimation accuracy.
\begin{equation}
    \text{RMSE}_{t} = \frac{\sum_{l=1}^{n_{x}+1}\sqrt{(\mathcal{Z}_{l,t}-f_{\text{est},l,t})^{2}}}{n_{x}+1}
\end{equation}
where $f_{\text{est},l,t}$ is a linear interpolation of the flow-front between the predicted flow-front progression along the measurement lines $i=1,...,n$ chosen from the range $l=1,...,n_{x}+1$.The rationale behind the choice of minimizing the one-step ahead prediction error between the estimated and the simulated flow-front is that, a correct prediction of the flow-front helps in the detection of any evolving heterogeneity in the flow-front. This information is necessary for the control of the production process and correction of any heterogeneity in real-time to avoid any moulding defects. 

The average RMSE values are shown in Fig. \ref{fig:data_table_plot9}-\ref{fig:data_table_plot11} to compare the estimation accuracy of the two models at different sampling rates, noise levels for the different number of line sensors and spatial approximation order included in the models. From the results it is seen that the higher order model improves the estimation accuracy for all combinations of sampling rates, noise levels and number of measurement lines.
From the results shown in Fig. \ref{fig:data_table_plot12} the average improvement of increasing the approximation order is calculated to be $16.7$\%. 

Furthermore, it is observed that with $4^{\textup{th}}$-order model, the estimation accuracy is higher for $8$ instead of $12$ line sensors. One of the possible reasons for this observation can be the over-fitting of the spatial domain caused by the higher order discretisation in combination with the higher order finite difference approximation of the spatial domain. Therefore, the SDE model with $8$ line sensors is considered the nominal model and is used for further evaluation.
\begin{figure}[!ht]
    \centering
    \begin{subfigure}[b]{0.45\textwidth} 
        \centering
        \includegraphics[width=0.75 \textwidth]{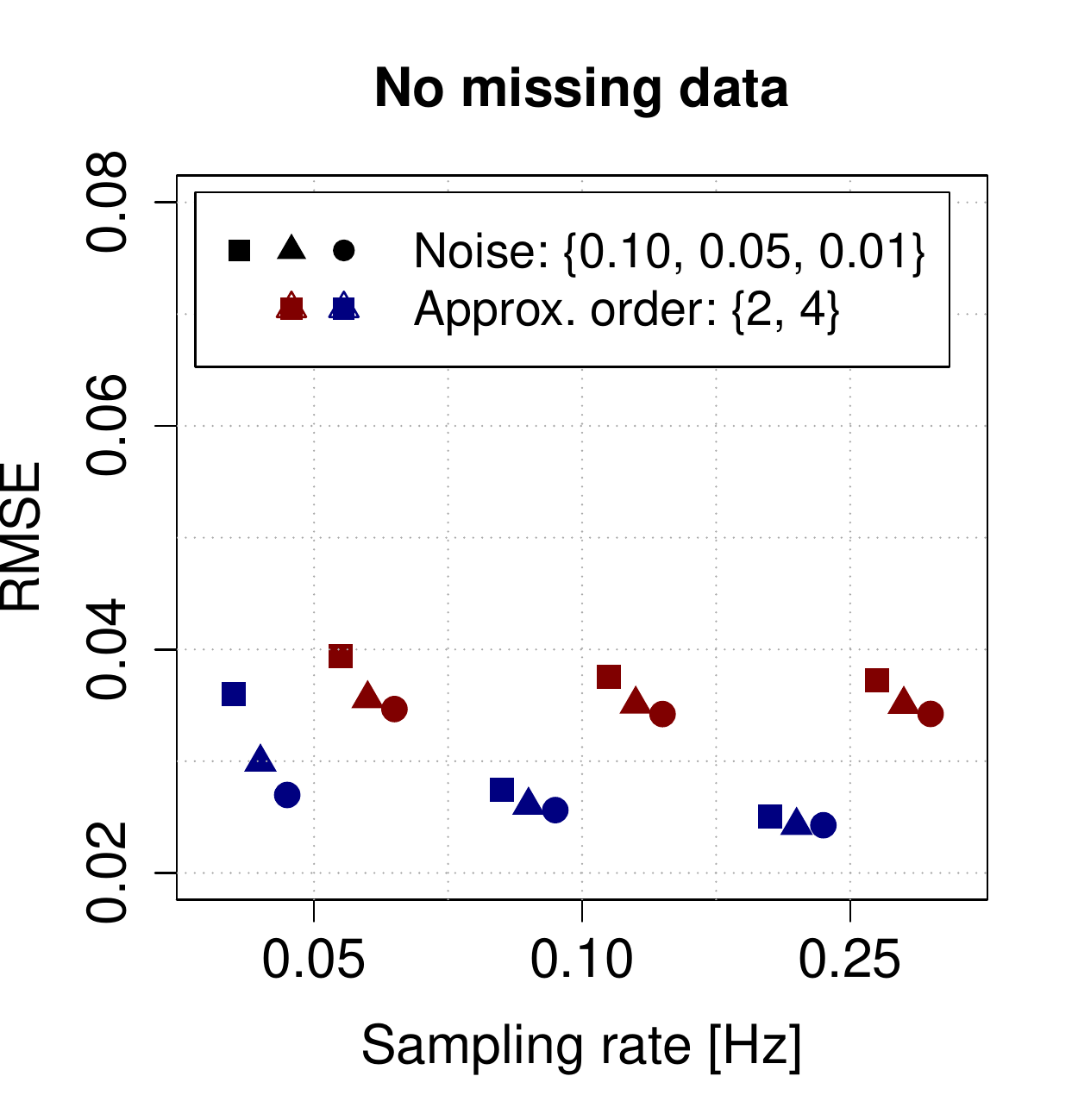}
        \caption{}
        \label{fig:rmse_4_2_order_plot1}
    \end{subfigure}
    \quad
    \begin{subfigure}[b]{0.45\textwidth} 
        \centering
        \includegraphics[width=0.75 \textwidth]{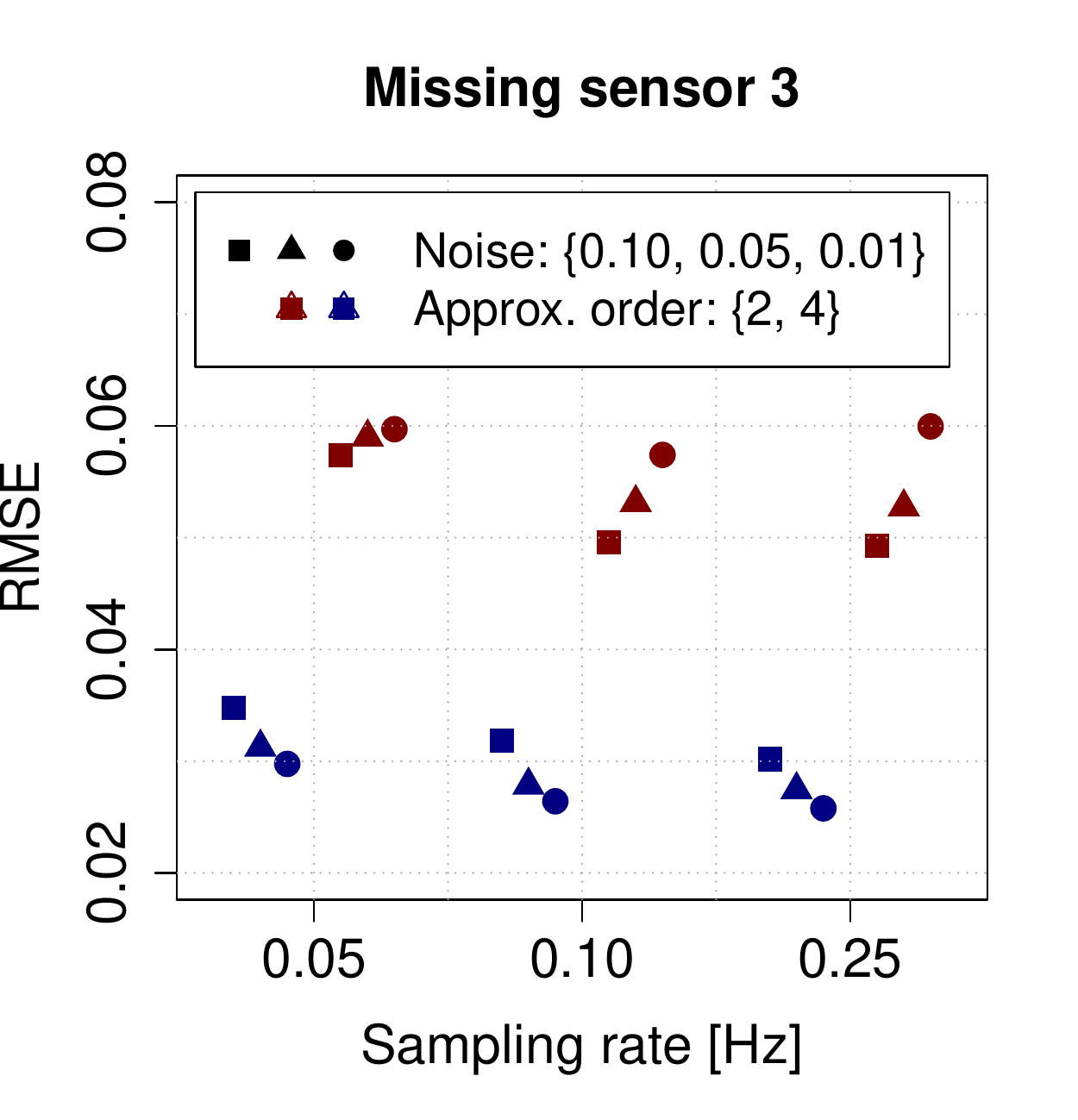}
        \caption{}
        \label{fig:rmse_4_2_order_plot2}
    \end{subfigure}
    \vskip\baselineskip
    \begin{subfigure}[b]{0.45\textwidth} 
        \centering
        \includegraphics[width=0.75 \textwidth]{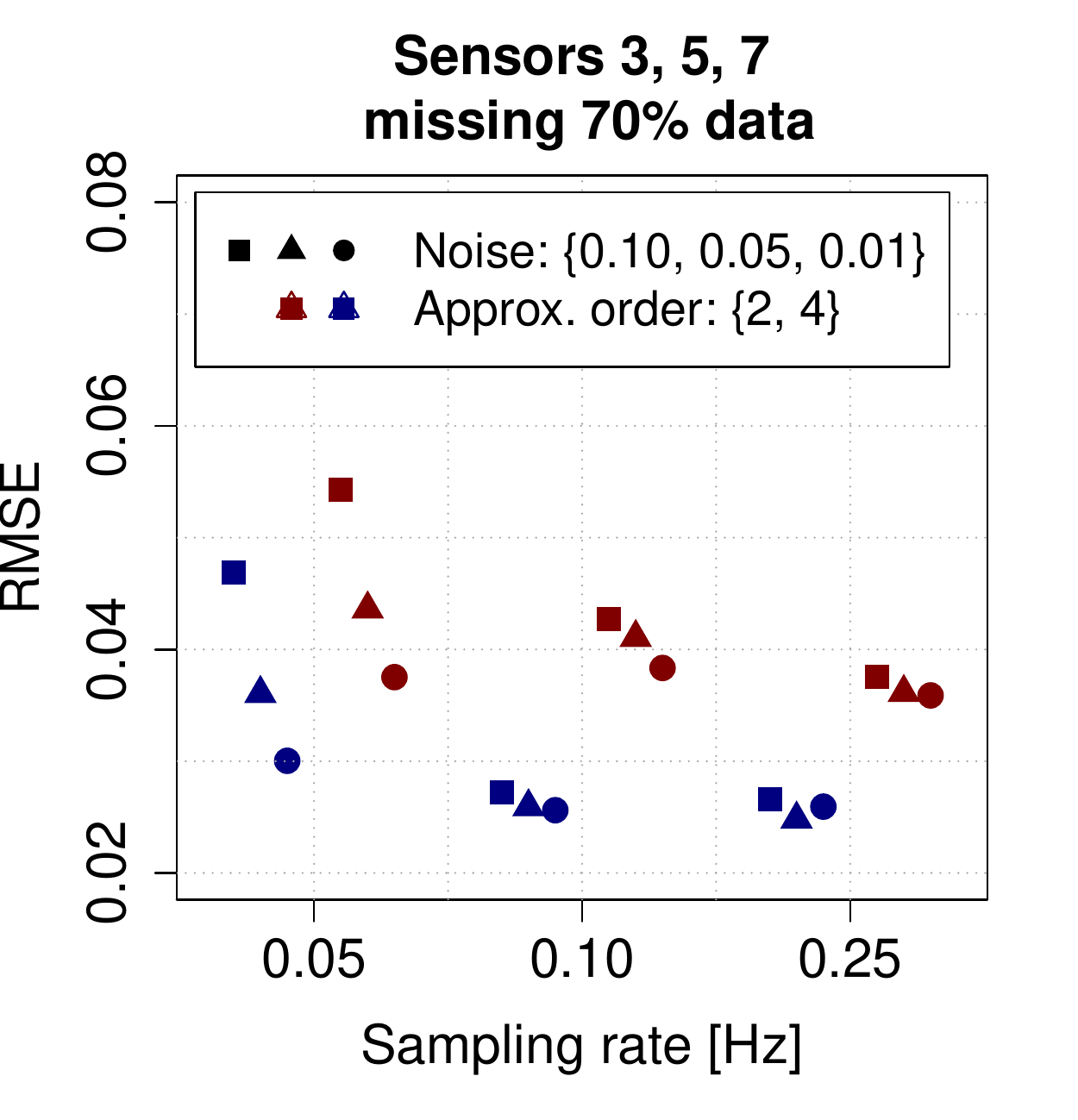}
        \caption{}
        \label{fig:rmse_4_2_order_plot3}
    \end{subfigure}
    \quad
    \begin{subfigure}[b]{0.45\textwidth} 
        \centering
        \includegraphics[width=0.75 \textwidth]{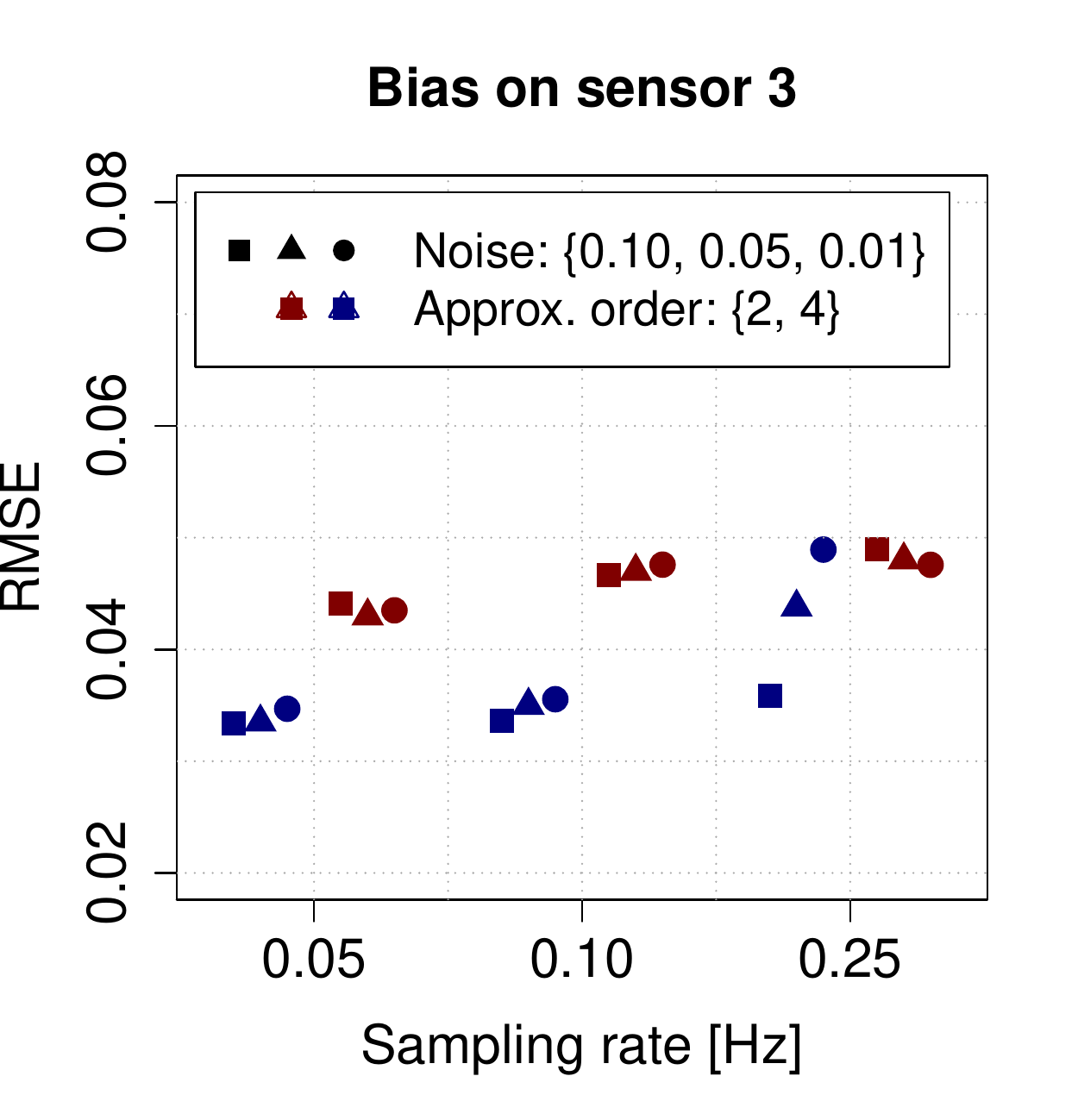}
        \caption{}
        \label{fig:rmse_4_2_order_plot4}
    \end{subfigure}
    \caption{The average RMSE values for the one-step ahead predictions of the flow-front using the models fitted to line sensor data for eight line sensors and using $2^{\textup{nd}}$ and $4^{\textup{th}}$-order finite difference approximation of the spatial domain. In (a) the RMSE values for a model estimated using full sensor information is shown for comparison. In (b), (c) and (d) the RMSE values for models estimated using data sets where sensor information was missing entirely, (b), or partially, (c), or where sensor data has been manipulated with a bias on 50 \% of the measurements. The data is perturbed around the specified sampling frequencies for presentation.}
    \label{fig:results_plots}
\end{figure} 
\subsubsection*{Case 1 - Missing information from one sensor}
In this case study, the influence of a complete sensor failure on the estimation accuracy of the flow-front dynamics is investigated. This is simulated by omitting the observation equation for line sensor $3$ from the observation model. From  Fig. \ref{fig:rmse_4_2_order_plot2}, it is observed that the estimation accuracy is almost unaffected for the $4^{\textup{th}}$-order  model, whereas the estimation accuracy of the $2^{\textup{nd}}$-order model decreases when sensor $3$ is missing.

\subsubsection*{Case 2 - Partially missing data from multiple sensors}
Here, the influence of several sensor faults during an infusion is investigated. It is simulated by randomly omitting $70$\% of the measurements from sensor $3$, $5$, and $7$. Two examples for sensor $3$ and $7$ are shown in Fig. \ref{fig:sensor3_7_miss}. From Fig. \ref{fig:rmse_4_2_order_plot3}, it is seen that the estimation accuracy of the $4^{\textup{th}}$-order model is mostly unaffected for the two highest sampling rates whereas for the $2^{\textup{nd}}$-order model, it is only true for the highest sampling rate. Furthermore, a decrease in the estimation accuracy is observed at the lowest sampling rate for the $4^{\textup{th}}$-order model.

\subsubsection*{Case 3 - Measurement bias on one sensor}
This case study investigates several sensor faults caused by additional potential shortcuts of a line sensor. It is simulated by introducing a bias with a value of $0.2$ meters on $50$\% of the measurements. Fig. \ref{fig:sensor3_bias} shows the implementation of this sensor bias. The results in Fig. \ref{fig:rmse_4_2_order_plot4} shows a decrease in the estimation accuracy both for the $2^{\textup{nd}}$ and $4^{\textup{th}}$-order finite difference approximation models. However, it is observed that the $4^{\textup{th}}$-order model with bias on the measurements generally estimates better than the $2^{\textup{nd}}$-order model without bias on the measurements.

The comparison between the two models for all case-studies are shown in Fig. \ref{fig:results_plots}. For all cases, the utilization of the $4^{\textup{th}}$-order finite difference approximation results in a better estimation accuracy at varying noise level and sampling rate. It is also seen that the $4^{\text{th}}$-order model with missing sensor data estimates more accurately than the $2^{\text{nd}}$-order model with no missing data. 

\begin{figure}[!ht]
    \centering
    \begin{subfigure}[b]{0.45\textwidth} 
        \centering
        \includegraphics[width=0.75 \textwidth]{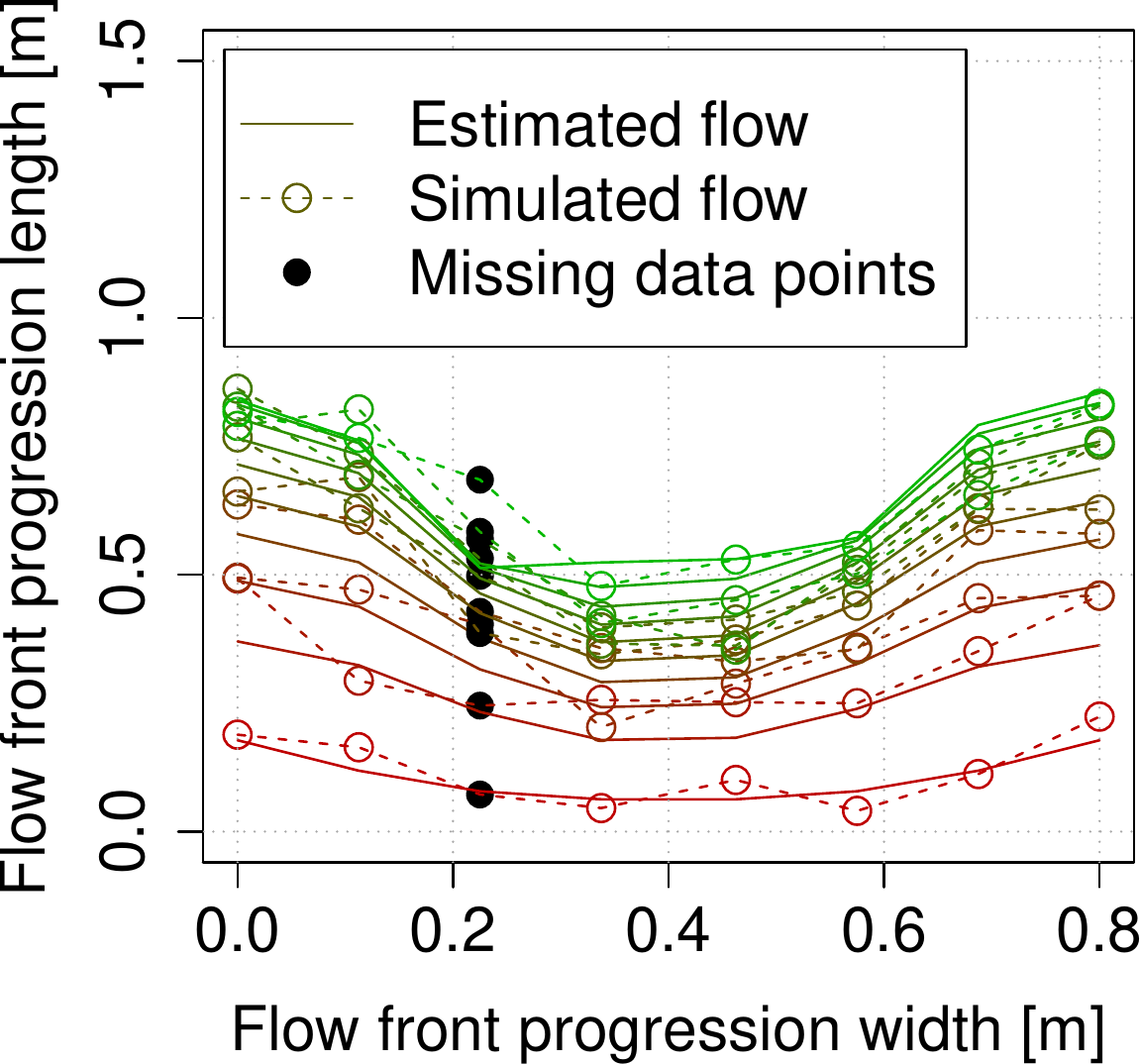}
        \caption{}
        \label{fig:missing_data_flow_front}
    \end{subfigure}
    \quad
    \begin{subfigure}[b]{0.45\textwidth} 
        \centering
        \includegraphics[width=0.75 \textwidth]{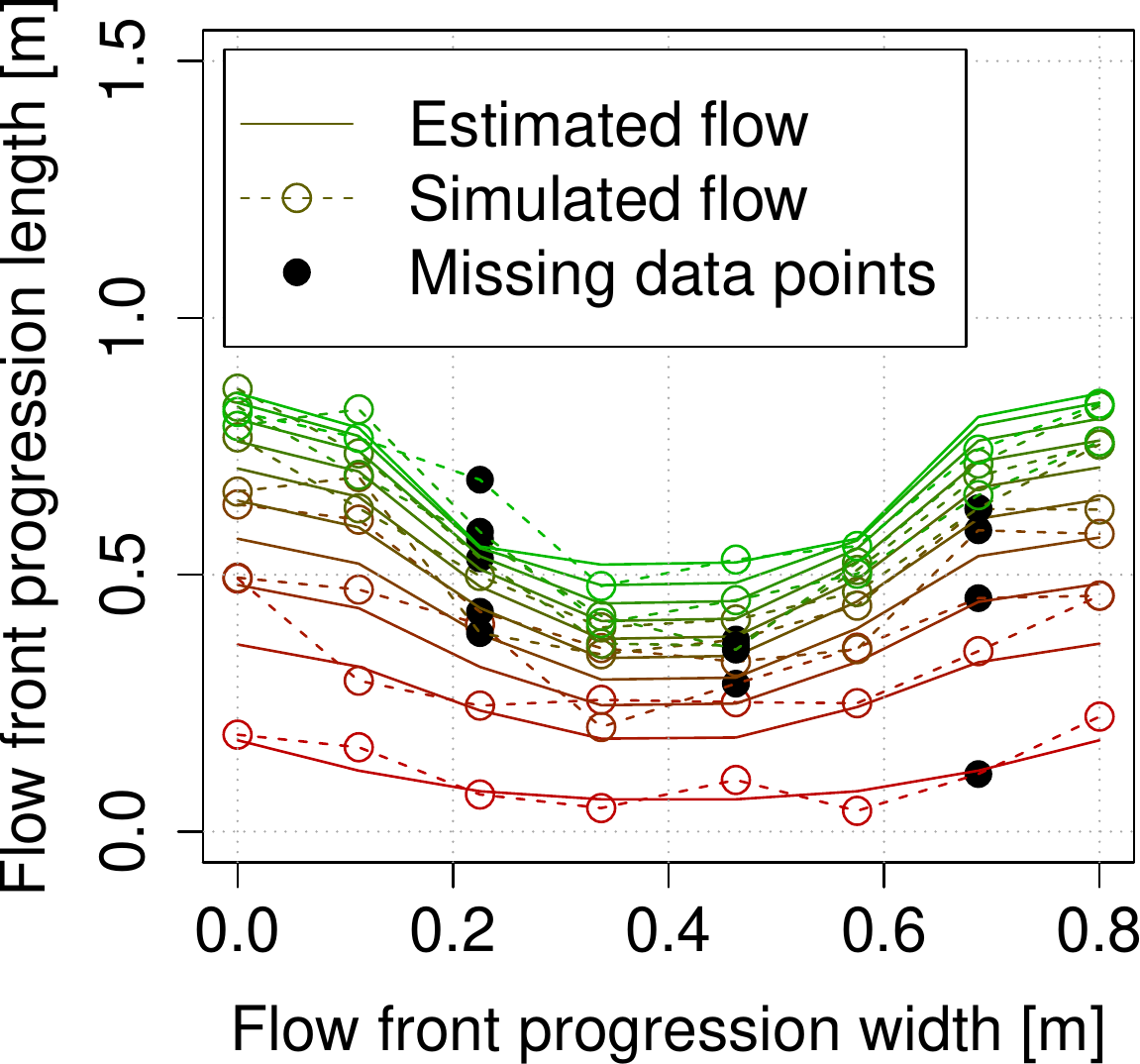}
        \caption{}
        \label{fig:missing_data_flow_front_357}
    \end{subfigure}
    \caption{Shows a sample plot of estimating the flow-front with missing information from sensor 3 in (a) and missing information from sensors 3, 5 and 7 in (b) respectively. The dot-dashed line indicates the simulated flow with added noise and the solid line shows the estimated flow-front. The line colour changes gradually with time from red to green.}
    \label{missing_data_flow_front_3_357}
\end{figure}
Fig. \ref{missing_data_flow_front_3_357} shows two examples of the estimated flow-front evolution with missing sensor information. It is clearly observed that the estimated flow-front trajectory follows the measured/simulated trajectory accurately even in the case of completely missing line sensor (see Fig. \ref{fig:missing_data_flow_front}) and also in the case of partially missing data from sensors 3, 5 and 7  (see Fig. \ref{fig:missing_data_flow_front_357}) respectively. 

\section{Conclusion}
\label{sec:conc}
Estimating the epoxy flow-front evolution in an environment with harsh conditions e.g. high temperature and strong chemicals increases the risk of sensor failures. This paper proposed a coupled SDE based flow-front modeling framework to handle missing observations during the flow-front progression. It uses the higher order finite difference approximation of the spatial domain and a CD-EKF estimation framework for SDEs which takes into consideration the effective dimension of the measurement space during the estimation. The proposed approach is shown to be effective in estimating the flow-front dynamics in multiple scenarios of sensor failures e.g. faulty, partially missing or fully missing line sensor data. In future, we will investigate the validity of such SDEs based virtual sensing framework using experimental faulty flow-front sensor data.  

\section*{Acknowledgment}
We thank the \MADE for the financial support and the colleagues at \SGRE facility in Aalborg, Denmark for their cooperation in problem formulation.

\bibliographystyle{elsarticle-num}
\bibliography{SDEMissingData}

\end{document}